\documentclass{aastex63}
\usepackage{tensor}

\shorttitle{}
\shortauthors{}

\newcommand{\Oi}{[O\,{\scriptsize I}]}
\newcommand{\Sii}{[S\,{\scriptsize II}]}

\newcommand{\OIa}{[O\,{\scriptsize I}]\,$\lambda$6300}
\newcommand{\OIb}{[O\,{\scriptsize I}]\,$\lambda$5577}

\newcommand{\SII}{[SII]$\lambda\lambda$6716,6731}
\newcommand{\SIIa}{[S\,{\scriptsize II}]\,$\lambda$6731}

\usepackage{soul}
\definecolor{mycolor}{RGB}{230,0,200}

\begin{document}

\title{Evidence for an MHD disk wind via optical forbidden line spectro-astrometry \footnote{Based on data collected by UVES (089.C-0299(A), 089.C-0299(B), P.I. I. Pascucci) observations at the VLT on Cerro Paranal (Chile) which is
operated by the European Southern Observatory (ESO).}}

\author{Whelan, E.T.}
\affiliation{Maynooth University Department of Experimental Physics, National University of Ireland Maynooth, Maynooth, Co. Kildare, Ireland}
\author{Pascucci, I.}
\affiliation{Lunar and Planetary Laboratory, The University of Arizona,
Tucson, AZ 85721, USA}
\author{Gorti, U}
\affiliation{SETI Institute/NASA Ames Research Center, Mail Stop 245-3, Moffett Field, CA 94035-1000, USA}
\author{Edwards, S}
\affiliation{Five College Astronomy Department, Smith College, Northampton, MA 01063, USA}
\author{Alexander, R.D.}
\affiliation{School of Physics \& Astronomy, University of Leicester, Leicester, LE1 7RH, UK}
\author{Sterzik, M.F.}
\affiliation{European Southern Observatory, Karl-Schwarzschild-Str. 2, 85748, Garching, Germany}
\author{Melo, C.}
\affiliation{European Southern Observatory, Casilla 19001, Santiago 19, Chile}


\begin{abstract}

{Spectro-astrometry is used to investigate the low velocity component (LVC) of the optical forbidden emission from the T Tauri stars RU~Lupi and AS~205~N. Both stars also have high velocity forbidden emission (HVC) which is tracing a jet. For AS~205~N, analysis reveals a complicated outflow system. For RU~Lupi, the \OIa\ and \SII\ LV narrow component (NC) is offset along the same position angle (PA) as the HVC but with a different velocity gradient than the jet, in that displacement  from the stellar position along the rotation axis is decreasing with increasing velocity. From the LVC NC PA and velocity gradient, it is inferred that the NC is tracing a wide angled MHD disk wind. A photoevaporative wind is ruled out. This is supported by a comparison with a previous spectro-astrometric study of the CO fundamental line. The decrease in offset with increasing velocity is interpreted as tracing an increase in the height of the wind with increasing disk radius. This is one of the first measurements of the spatial extent of the forbidden emission line LVC NC ($\sim$ 40~au, 8~au for RU~Lupi in the \SIIa\ and \OIa\ lines) and the first direct confirmation that the LVC narrow component can trace an MHD disk wind.} 
\end{abstract}


\section{Introduction} 
A crucial step in understanding how stars accrete their mass, as well as how disks evolve, is clarifying how
the accreting disk gas loses angular momentum. Originally MHD disk winds were favored \citep{Koenigl1993} but the prevailing view for several decades has been that MRI-induced turbulence \citep{Balbus1991} leads to outward viscous transport of angular momentum \citep{Armitage2011}, enabling disk material to flow radially inward. However, recent simulations find that non-ideal MHD effects suppress MRI over a large range of disk radii ($\sim 1-30$\,au), restoring radially extended MHD disk winds as the prime means for
extracting angular momentum and enabling accretion at the observed rates \citep{Bai2013, Gressel2015}.

On the observational side, there has been renewed interest in identifying disk wind tracers and testing the emerging paradigm of disk evolution. See \cite{EP2017} for a recent review. Emission from optical forbidden lines has been a long-established tracer of flowing material from young stars \citep{Edwards1987} and recent observations have helped clarify its origin.  The so-called ``high-velocity” component (HVC), emission blueshifted by $>$\,30\,km/s \citep{Hartigan1995}, is confirmed to be associated with collimated micro-jets (e.g., Hartigan \& Morse 2007) and its \OIa\ luminosity is found to be better correlated with accretion luminosity than with stellar luminosity or mass \citep{Nisini2018}. High-resolution ($ \Delta v < 10$\,km/s) spectroscopy revealed that the lower velocity component (LVC) can be often described  as the combination of two Gaussian profiles (a “broad component,” BC, and a “narrow
component,” NC; \citep{Rigliaco2013, Simon2016, McGinnis2018}. 
The LVC-BC blueshifts and large FWHMs point to an MHD disk wind origin \citep{Simon2016} while the origin of the LVC-NC is less certain. However, two observables suggest that the NC most likely traces the same MHD wind, instead of a photoevaporative thermal wind\citep{Weber2020}: i)  its peak centroids
and FWHMs correlate with those of the BC \citep{Banzatti2019} and ; ii) the observed \SIIa{} LVC luminosities are much weaker than predicted in photoevaporative winds \citep{Pascucci2020}. Clarifying whether the NC truly traces an MHD disk wind and measuring its vertical extent is critical to estimate wind mass loss rates (Fang et al. 2018) and directly test one of the main predictions of the disk wind theory, i.e., that wind mass loss rates are similar to mass accretion rates.

Here, we apply the spectro-astrometry (SA) technique to the optical forbidden lines from RU~Lupi and AS~205~N, two well known late K-type stars with evidence for jets and winds (see Section~\ref{sect:targets} and Table~\ref{propsample} for further details). Both of our targets are among a handful of TTS with high accretion rates, that also show strong \OIa\ with very broad high velocity emission that is blended with the low velocity emission, and a low velocity component that includes both a NC and a broad BC component, where the NC dominates over the BC. Moreover, in contrast to the majority of accreting TTS where the LVC is not detected in \SIIa, our two targets show prominent LVC emission at \SIIa\ as well as \OIa. RU~Lupi and AS~205~N have previously been investigated with SA \citep{Takami2001, Takami2003} and the results of these works are discussed in Sections 2.1 and 3. While the results presented here confirm the previous results, the much improved spectral resolution and positional uncertainty of our study, allowed new details to be revealed about the kinematics and spatial properties of the line regions under investigation. Furthermore, \cite{Takami2003} did not investigate the FELs of the AS~205 system but analysed several permitted lines (including H$\alpha$) to search for a binary signal.

SA is a powerful technique for recovering spatial information from spectra below the angular resolution limitations of the observations \citep{Bailey1998, Whelan2008, Cahill2019}. Therefore, our goal is to use this technique to explore the origin of the low velocity forbidden emission in both stars. The technique is based on measuring the centroid of the spatial
profile of an emission line region as a function of wavelength,
and with respect to the centroid of the spatial profile of the continuum. Frequently, this has been
done using Gaussian fitting to produce what is referred to as a
position spectrum. For ground-based spectroscopy the accuracy to which this can be done is dependent on the signal to noise ratio (SNR) of the observation and the seeing, and sub-milliarcsecond precision has been achieved at near-infrared wavelengths with for example VLT/CRIRES \citep{Pontoppidan2011}. This study focuses on the \OIb, \OIa\ and \SIIa\ lines and the work is organised as follows. In Section~\ref{obs} the targets, observations and reduction of the data are described. In Section 3 the spectra and spectro-astrometric results are discussed. Sections 4 and 5 are devoted to the Discussion and Summary.

\begin{figure*}
\centering
\includegraphics[width=18cm, trim= 0cm 0cm 0cm 0cm, clip=true]{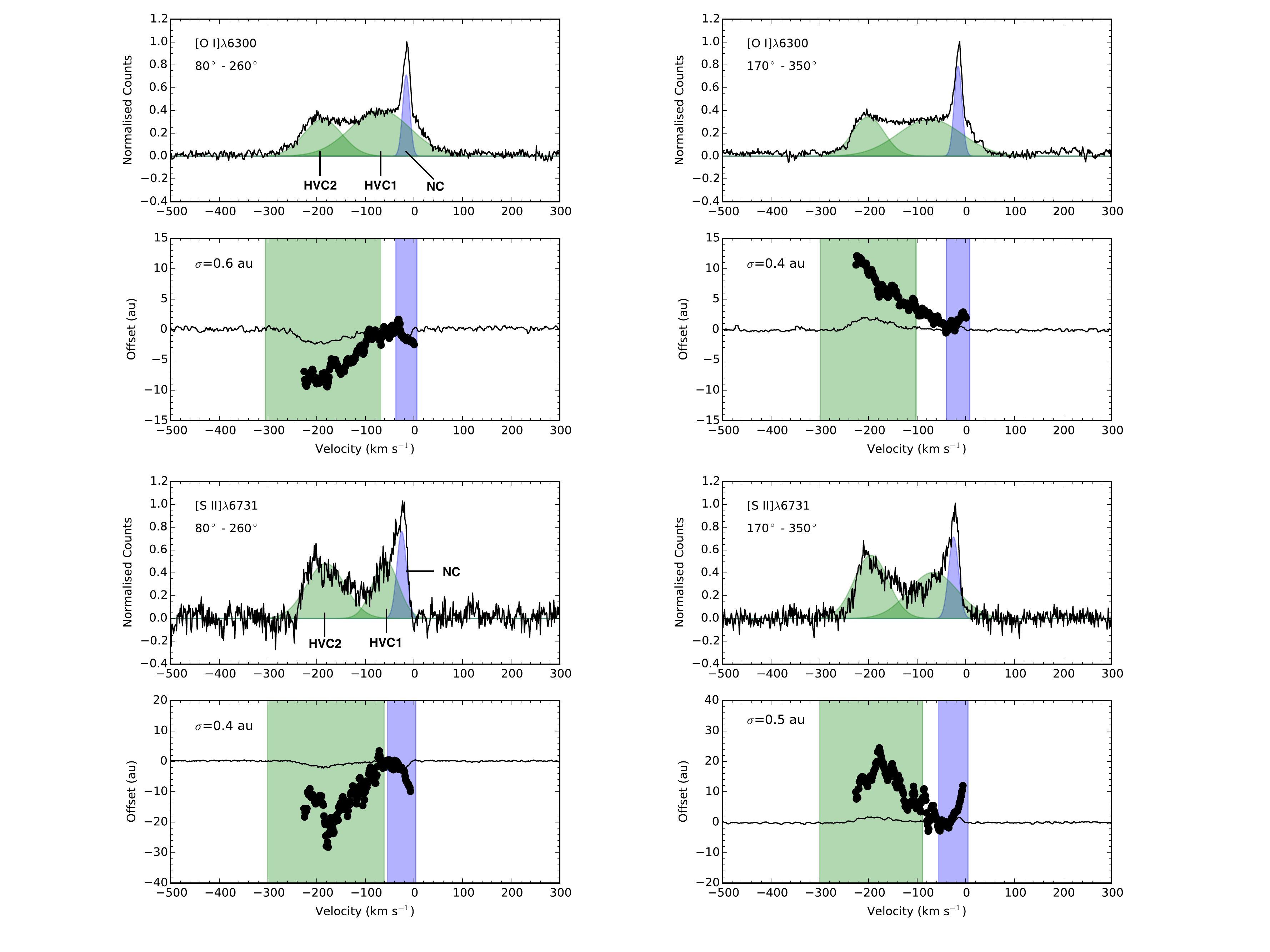}
   \caption{{\bf RU Lupi:} Spectro-astrometric analysis of the [O I]$\lambda$6300, and [S II]$\lambda$6731 lines. Top: The continuum subtracted line emission, normalised to the line peak. The fitted kinematic components are over-plotted with with the LVC-NC shown in blue, and the HVCs in green. Bottom: The position spectra with the ranges of the LVC-NC and HVC shown as the colored shaded regions. The continuum subtracted offsets are over-plotted in black.}
  \label{RULUP_SA}     
\end{figure*}

 \begin{figure*}
\centering
   \includegraphics[width=16cm, trim= 0cm 3cm 0cm 3cm, clip=true]{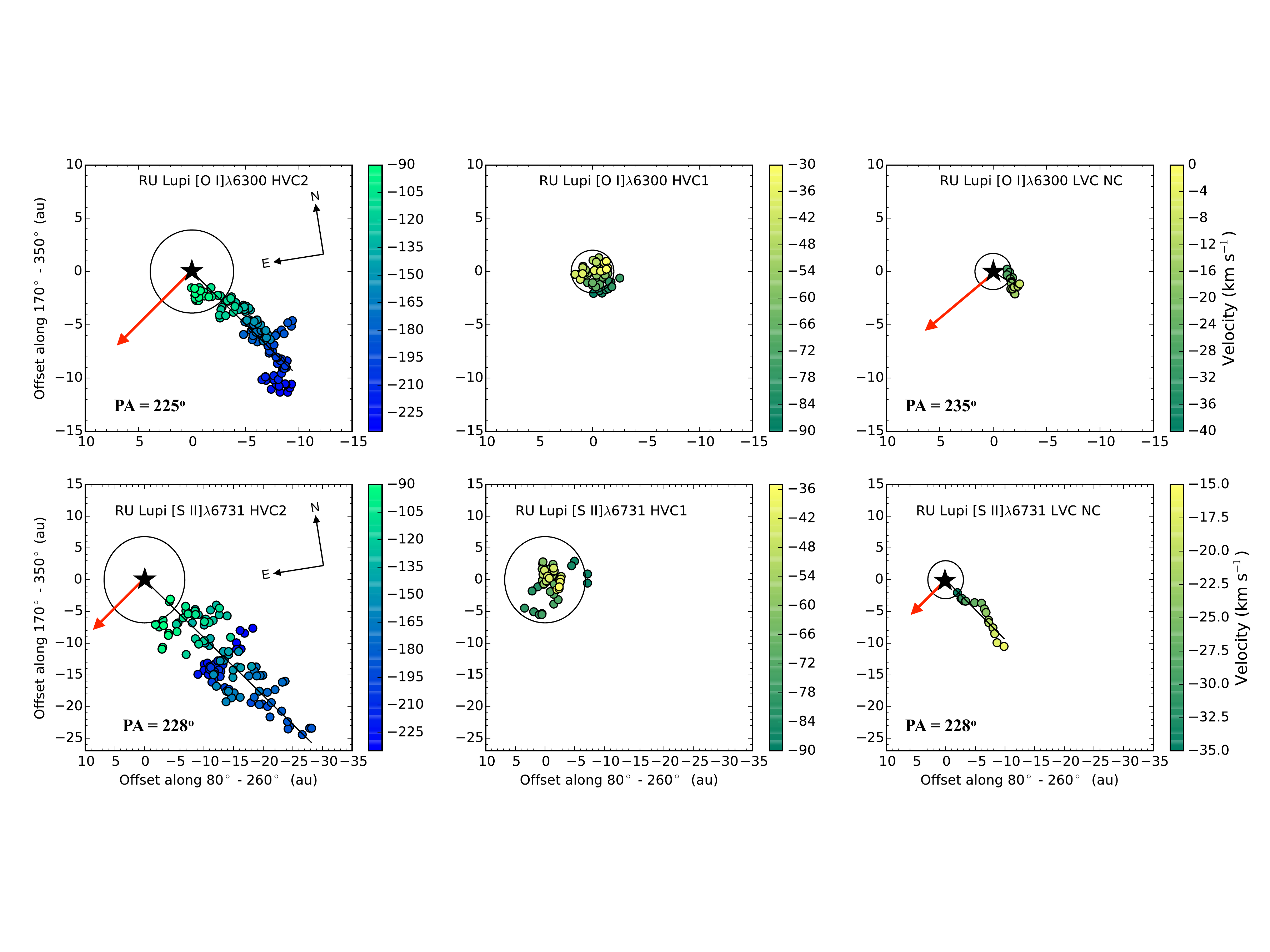}
     \caption{{\bf RU~Lupi:} Combining the perpendicular slit PAs to recover the 2D centroid at each wavelength(velocity). The continuum emission has been subtracted here to reveal the full displacement in the outflow. The black circle is the 3-$\sigma$ uncertainty in the centroid, the black line is the best fit PA to the flows and the red arrow is the PA of the RU Lupi disk major axis.}
  \label{RULUP_CONT_2D}     
\end{figure*}

\section{Targets, Observations and Data Reduction}\label{obs}

\subsection{Targets}\label{sect:targets}
RU~Lupi is a CTTS located in the Lupus Star forming region at d $\approx$ 140~pc. \cite{Takami2001} used spectro-astrometry to investigate the H$\alpha$ emission of RU~Lupi, and found that there is a distinct spatial symmetry in the offset for blue and red shifted wings of the H$\alpha$ line. This signal was attributed to a bipolar outflow driven by RU~Lupi and the bipolar outflow theory was supported by their spectro-astrometric study of the \OIa\ and \SIIa\ lines. The PA of the outflow was estimated at around 225$^{\circ}$ from this work. \cite{Pontoppidan2011} measure a spectro-astrometric signal in the CO fundamental line consistent with that of a wide-angle molecular wind. The DSHARP study measured the PA of the RU~Lupi disk at $\sim$ 121$^{\circ}$ \citep{Huang2018}, which is approximately perpendicular to the outflow direction estimated by \cite{Takami2001}. 

AS~205 is a triple system (AS~205~N, AS~205~Sa,b) \citep{Kurtovic2018}, located in the northern region of the Ophiuchus star-forming region at d $\sim$ 127~pc \citep{GAIA2018}. It's primary star AS~205~N (also AS~205~A) is a CTTS and it is distinguished by its bright molecular emission \citep{Salyk2008}. AS~205~N and AS~205~S have been detected at a projected separation of 1\farcs3 and a PA of $\sim$ 212$^{\circ}$ \citep{McCabe2006}. The DSHARP study measures disk PAs of 114$^{\circ}$~$\pm$12$^{\circ}$ and 110$^{\circ}$~$\pm$2$^{\circ}$ for AS~205~N and AS~205~S respectively \citep{Kurtovic2018}. 

\cite{Pontoppidan2011}  found evidence for  a molecular wind using spectro-astrometry in the CO fundamental line. \cite{Takami2003} carried out a spectro-astrometric study of the H$\alpha$ emission from the AS~205 system. Analysis showed a displacement along a PA of $\sim$ 33$^{\circ}$ (E of N) which is interpreted as a binary signature. The authors suggest that any difference between the PA of the displacement and the PA of AS~205~N and AS~205~S can be explained by the orbital motion of the system.

\begin{figure*}
\centering
\includegraphics[width=18cm, trim= 0cm 0cm 0cm 0cm, clip=true]{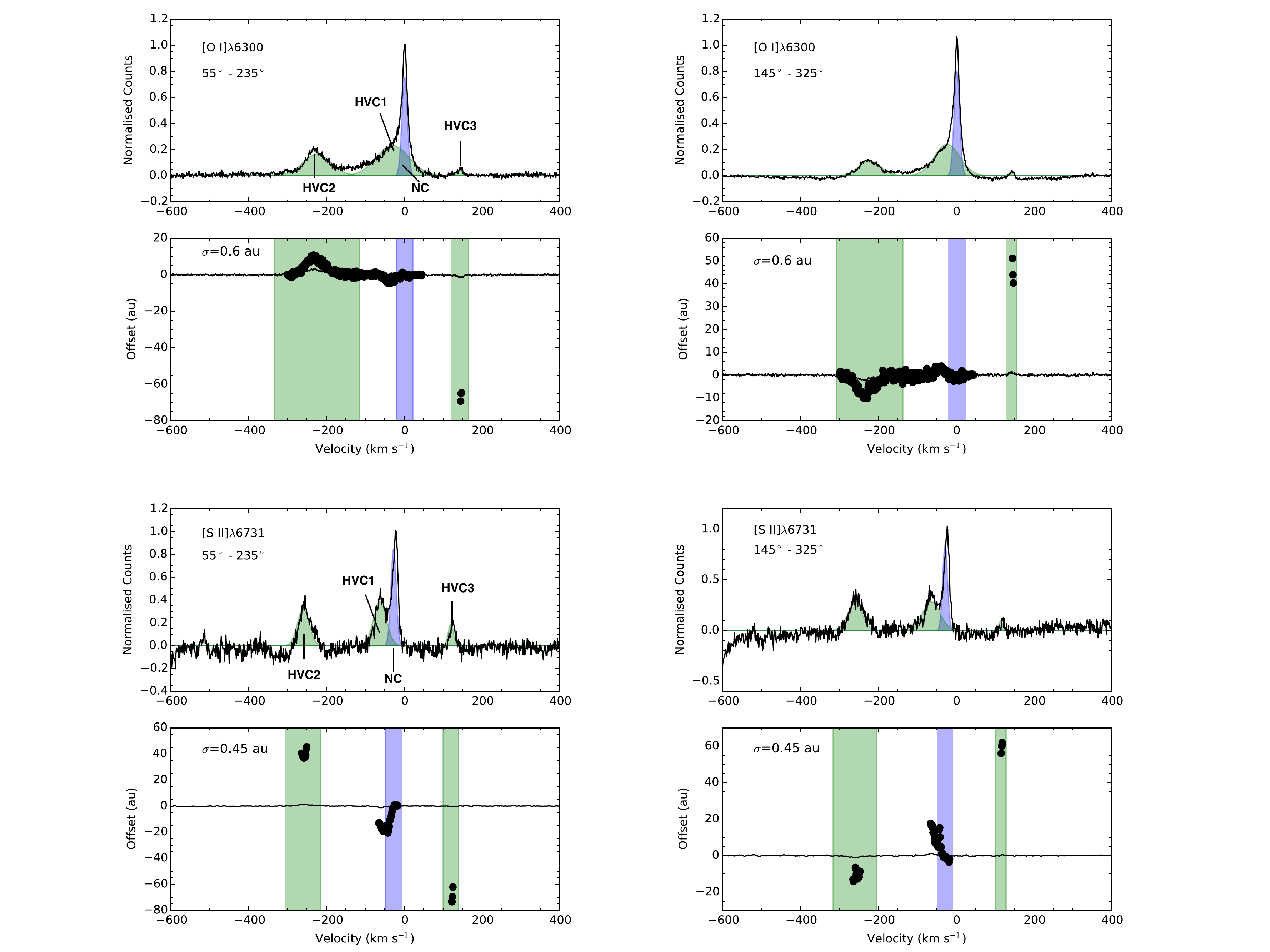}
  \caption{{\bf AS 205:} Spectro-astrometric analysis of the [O I]$\lambda$6300, and [S II]$\lambda$6731 lines. Top: The continuum subtracted line emission, normalised to the line peak. The fitted kinematic components are over-plotted with with the LVC-NC shown in blue and the HVCs in green. Bottom: The position spectra with the ranges of the LVC-NC and HVCs shown as the colored shaded regions. The continuum subtracted offsets are over-plotted in black.}
  \label{RULUP_full}     
\end{figure*}

\begin{figure*}
\centering
   \includegraphics[width=19cm, trim= 0cm 5cm 0cm 5cm, clip=true]{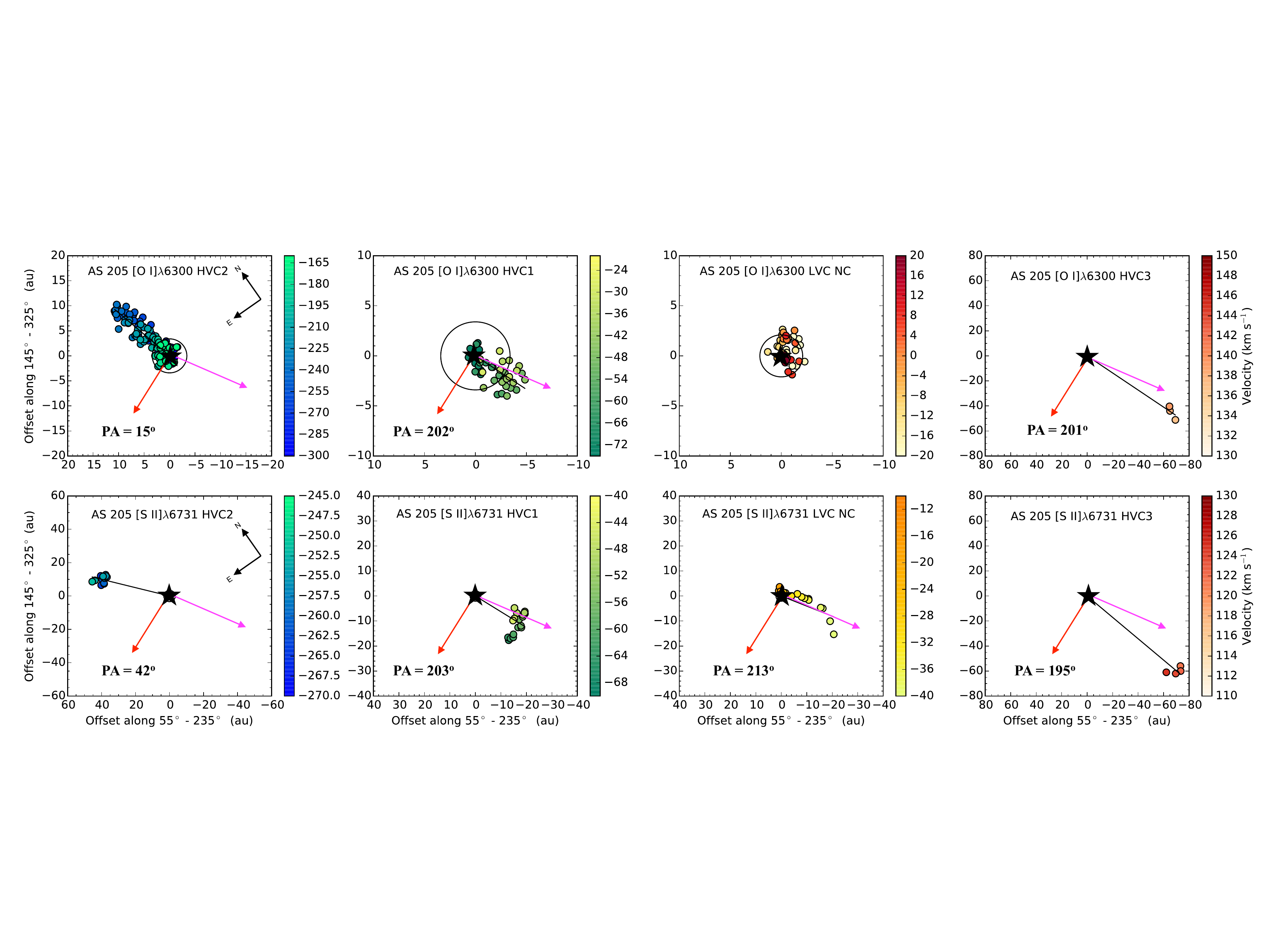}
     \caption{{\bf AS~205~N:} Combining the perpendicular slit PAs to recover the 2D centroid at each wavelength(velocity). The continuum emission has been subtracted here to reveal the full offset in the outflow. The black circle is the 3-$\sigma$ uncertainity in the centroid, the red arrow the disk PA, the magenta arrow the binary PA and the black arrow the best PA fit to the signals. These results are difficult to explain if the triple system is driving only one outflow.}
  \label{AS205N_CONT_2D}     
\end{figure*}


\subsection{Observations and Data Reduction}\label{sect:obs}

Spectra of RU Lupi and AS~205 were collected with the UV-Visual Echelle Spectrograph (UVES) on the European Southern Observatory's Very Large Telescope (ESO VLT) in May and June 2012 \citep{Dekker00} through the ESO programme 089.C-0299 (PI. I. Pascucci). The wavelength range was approximately 5000~\AA\ to 7000~\AA{} and a spectral resolution of R = 40,000 was achieved. As the spectro-astrometric precision is given by the following equation 

\begin{equation}
\sigma = \frac{FWHM}{2.35\sqrt{N_{p}}}   
\end{equation}
where N$_{p}$ is the number of detected photons and the FWHM is the seeing, the exposure times were set, assuming typical seeing, to reach 2~mas precision. This goal was achieved and sub 2~mas precision was reached, in general, for the emission line regions. Additional details on the observations are given in Table~\ref{sample}.

The slit width was set to 1\arcsec{}. With UVES it is possible to rotate the slit PA to any position on the sky. For both of our targets, spectra were obtained at four orthogonal slit PAs. The slit PAs were chosen so that spectra were obtained parallel and anti-parallel to the known accretion disk PA\footnote{Note that the disks PAs were subsequently updated by the DSHARP study. This is the reason for the difference between the disk PAs given in Table 1 and the slit PAs in Table 2.} and parallel and anti-parallel to any outflow emission. It was assumed that the PA of any outflow emission is perpendicular to the disk PA. 
The strategy of obtaining four perpendicular slit PAs is integral to the spectro-astrometric analysis. Firstly, the observations at anti-parallel slit PAs allow for spectro-astrometric artefacts to be ruled out \citep{Whelan2015, Brannigan2006}. Secondly, spatial offsets measured in spectra taken at perpendicular slit PAs can be combined to recover the on-sky PA of the extension in the emission line region.
Further details on the observations are given in Table 2.

The data were initially reduced using the ESO pipeline. \cite{Whelan2015} discuss how the re-binning of X-Shooter spectra by the ESO pipeline can introduce a false spectro-astrometric signal due to a spatial aliasing effect. A similar problem was found for some of the observations, when the 2D spectra produced by the UVES pipeline were analysed using SA. For this reason the spectra were also reduced using standard IRAF routines for the reduction of echelle spectra and it was this data that was used in the analysis and that is presented below. It was found by the authors that using IRAF for the data reduction reduced the spatial aliasing problem. The SA analysis was done using standard fitting routines. The continuum subtraction was done using the IRAF {\it continuum} routine. With this routine the continuum is fitted and removed from the spectrum row by row, where a row is along the dispersion direction \citep{Whelan2008}. For each row only the continuum emission is fitted by masking the emission line regions. The spectra were also also corrected for telluric and photospheric lines using the method outlined in \cite{Hartigan1995} and adopted in subsequent papers \citep[e.g.,][]{Rigliaco2013, Banzatti2019}.

\begin{deluxetable*}{lccccccc}
\tablecaption{Spectro-astrometric sample: properties relevant to this study \label{propsample}}
\tablehead{
\colhead{Object} & \colhead{SpTy} & \colhead{Log($\dot{M}_{\rm acc}$)} & \colhead{Disk} & \colhead{PA\tablenotemark{*}} & Disk Inc  &\colhead{Wind}  & \colhead{References} \\
            &     & \colhead{[M$_\odot$/yr]} & \colhead{Type} & \colhead{[$^\circ$]} & \colhead{[$^\circ$]}  &\colhead{Diagnostic} & 
            }
\startdata
RU~Lup  & K7 & -7.77 &  full & 121 &19 & CO, \Oi  & Hu18, C11,P11,F18 \\
AS~205~N & K5 & -6.58 & full & 114 &20 & CO, \Oi   & K18, F18,P11  \\
AS~205~S &K7 & - & - &110 &66 &-   &K18  \\
\enddata
\tablenotetext{*}{This is the position angle of the disk major axis. The most recent values are given here and therefore in some cases they differ from the slit PAs chosen to be parallel to the disk PAs (see section 2.3 and Table 4.)}
\tablecomments{References stand for: C11 = Curran et al. (2011); F18 = Fang et al. (2018); Hu18 = Huang et al. (2018); P11 = Pontoppidan et al. (2011); K18= Kurtovic et al. (2018)} 
\end{deluxetable*}


\begin{deluxetable*}{ccccccc} 
\tablecaption{Observing log \label{sample}}
\tablehead{
\colhead{Object} &   \colhead{RA (J2000)}  &   \colhead{Dec (J2000)}  & \colhead{Date} & \colhead{Slit PA} & \colhead{Exposure Times} & \colhead{Seeing} \\
&   \colhead{(J2000)}  &   \colhead{(J2000)}  & \colhead{(dd-mm-yyyy)} & \colhead{($^\circ$)} & \colhead{(ncycle$\times$s)} 
 & \colhead{(arcsec)}}
\startdata
RU Lupi   &   15:56:42.3  &-37:49:15.5 &29-06-2012 &80 &5 $\times$ 300 &1.3 \\
                 &   & &28/29-06-2012 &170 &10 $\times$ 300 &1.2 \\
                 &   & &29-06-2012 &260 &5 $\times$ 300 &1.0 \\
                 &   & &28/29-06-2012 &350 &10 $\times$ 300 &1.4 \\
AS 205      & 16:11:31.4 & -18:38:28.3 &02-05-2012 &55 &6 $\times$ 300 &0.77 \\
                     &  & &02-05-2012 &145 &6 $\times$ 300 &0.76 \\
                    &  & &02-05-2012 &235 &6 $\times$ 300 &0.8 \\
                    &  & &02-05-2012 &325 &6 $\times$ 300 &0.68 \\
\enddata
\end{deluxetable*}

\section{Results}

The UVES spectra cover two oxygen forbidden lines, one at 6300\,\AA, and the other at 5577\,\AA , with the latter being the weaker of the two (e.g., Simon et al. 2016), and the sulphur forbidden line doublet at 6716\,\AA, 6731\,\AA. The oxygen profiles from both sources, obtained at a resolution similar to UVES, were previously analyzed in \citet{Fang2018,Banzatti2019}.  These works separated CTTS line profiles into distinct kinematic components by finding the minimum number of Gaussian profiles that reproduce the observed one. Components are identified as HVC, where the absolute value of the Gaussian velocity centroid is larger than 30\,km~s$^{-1}$, or LVC, with centroids within 30\,km~s$^{-1}$ of the photospheric velocity. The LVC can be further decomposed into a BC or NC (FWHM larger  or smaller than 40\,kms$^{-1}$), see e.g. \cite{Simon2016}. In those studies both RU~Lupi and AS~205~N were found to have quite broad HVC emission,  requiring fits with more than one component, and their LVC showed both BC and NC. 

 We follow a similar approach here.   Based on this approach, both \Oi{} lines of RU~Lupi and AS~205~N have an LVC and two (RU~Lupi) or three (AS~205~N) HVC components in the \OIa{} transition. For both stars only a LVC NC is identified. A similar decomposition scheme is applied to the [S II] lines (see Table \ref{fits})\footnote{Where the \OIa{} and \SIIa{} lines are detected the same profiles are seen in the [O I] $\lambda$ 6363 and [S II]$\lambda$ 6716 lines, albeit at a lower SNR.}. The fits to the \OIa\ and \SIIa\ lines are shown in Figures 1 and 3 and the kinematic results presented in Table \ref{fits}. All velocities quoted here are in the stellocentric reference frame. The differences between the results of the kinematical fitting presented here, and those of previous studies, for the \OIa\ line, are explained by variability in the HVC and this is discussed in detail in the appendix.

As described in Section~\ref{sect:targets} the AS~205 system is a multiple system with three known members. The aim here was to investigate the emission from AS~205~N. However, for the 55$^{\circ}$ and 235$^{\circ}$ slit PAs AS~205~S~a,b are also included in the UVES slit. Therefore, given a separation of 1\farcs3 between the N and S components, and a seeing of $\sim$ 0\farcs8, the emission from AS~205~S could also be evaluated at these PAs (although AS~205~S a,b are not resolved). For AS~205~S only the \OIa\ line was detected, showing a weak LVC BC. The spectro-astrometric analysis for AS~205~S will be discussed in a separate paper. As the binary PA is $\sim$ 212$^{\circ}$ and the slit width was 1~\arcsec, there was no emission from AS~205~S in the 145$^{\circ}$ and 325$^{\circ}$ spectra.



Next, we will be presenting the spectro-astrometric results and the following points should be noted for all the results.

\begin{enumerate}
      \item The method of \cite{Takami2001} where the position spectra for the anti-parallel slit positions were subtracted was followed. \cite{Takami2001} discuss how this technique can remove artefacts caused by for example, optical distortion introduced by the spectrograph, misalignment of the spectrum within the CCD columns, or imperfect flat fielding. 
    \item The 1-$\sigma$ uncertainties in the centroid position given in Figures \ref{RULUP_SA} and \ref{RULUP_full} are the average values for the whole spectral range analysed. The precision will increase at the emission line peaks due to an increase in the SNR. 
    \item The continuum contamination was removed using the continuum subtraction method outlined in Section 2.2. Position spectra before and after continuum subtraction are shown.  
    \item The spectro-astrometric analysis shown in Figures \ref{RULUP_CONT_2D} and \ref{AS205N_CONT_2D} does not provide any information on the width (disk radial extent) of the flow, as the spectro-astrometric centroid measures only the the position of the brightest part of the emission region. Any spread perpendicular to the jet PA (black line), is related to the accuracy to which the emission centroid is measured. Spectra taken perpendicular to the jet axis would be required to recover the flow width.
    \end{enumerate}

\begin{deluxetable*}{llllll}
\tablecaption{Classification of the \OIb, \OIa, and \SIIa\ lines.} \label{fits}
\tablehead{ 
\colhead{Object} &\colhead{Line}  &\colhead{LVC NC} &\colhead{LVC BC}   &\colhead{HVC Blue} &\colhead{HVC Red} \\ 
& \colhead{}  &\colhead{\small V$_{rad}$, FWHM (km~s$^{-1}$)} &\colhead{ \small V$_{rad}$, FWHM (km~s$^{-1}$)} &\colhead{\small V$_{rad}$, FWHM (km~s$^{-1}$)} &\colhead{\small V$_{rad}$, FWHM (km~s$^{-1}$)}
}
\startdata
RU~Lup &\OIa   &-13, 18.2 & &-66.11, 149.9 ({\bf HVC1}) & \\
 & & & &-197.5, 84.5 ({\bf HVC2})  & \\
& \SIIa  &-23.9, 26.6 &  & -100.1, 157.1 ({\bf HVC1}) &\\
& & & &-194.71, 59.5 ({\bf HVC2}) & \\ 
AS~205~N & \OIa  &-2.35, 16.8 &  &-30.3, 88.2 ({\bf HVC1}) &139.5, 13.7 ({\bf HVC3}) \\
& & & &-226.5, 76.4 ({\bf HVC2})   \\
& \SIIa  &-27.3, 16.3 &  &-62.3, 39.8 ({\bf HVC1}) &119.8, 16.1 ({\bf HVC3}) \\ 
& & & &-259.2 , 35.6 ({\bf HVC2}) \\
AS~205~S &\OIa &    &-2.6, 42.9 & &  \\ 
\enddata
\end{deluxetable*}

\subsection{RU Lupi: a vertically extended LVC-NC}
For RU Lupi only the blue-shifted outflow is detected in forbidden emission, and vertical offsets from the star-disk plane are measured in the \OIa\  and [S II]$\lambda$6731, $\lambda$6716  lines (see Figure \ref{RULUP_SA} where the \OIa\ and \SIIa\ results are shown). No spectro-astrometric signal is measured in the highest critical density ($\sim 10^{8}$\,cm$^{-3}$) \OIb\ line. The 1-$\sigma$ uncertainty in the SA can be taken as an upper limit on the extent of the \OIb\ emission region. For RU~Lupi this is 3~au (after continuum subtraction) which equals 9~au when the inclination of the system is accounted for. For both the \Oi{} and \Sii{} lines it is found that the LVC NC  and the HVC2 are displaced from the stellar position while the HVC1 is coincident with the stellar position (within the uncertainty of the centroid fitting). Note that the \SIIa\ emission is more extended than the \OIa\ emission, for both the LVC and HVC2, as would be expected due to the lower critical density of the \SIIa\ line ($\sim 2 \times 10^4$\,cm$^{-3}$) with respect to the \OIa\ ($\sim 2 \times 10^6$\,cm$^{-3}$). The PA of the extended emission can be recovered by combining the results after continuum subtraction for the perpendicular slit positions. This is shown in Figure 2 with a black line. The LVC NC, HVC1, and HVC2 are shown separately. It is clear that the HVC1 is not extended, and we suggest that it is tracing the outflow close to it launching point. An average PA of 229$^{\circ}$ $\pm$ 10$^{\circ}$ is measured for the extended emission in agreement with the spectro-astrometric analysis of \cite{Takami2001}. These results reveal two important properties of the LVC NC. Firstly, the LVC NC is extended along the same PA as the HVC2 for both the \OIa\ and \SIIa\ lines. This rules out its origin in the disk and connects it to the jet. Secondly, the velocity gradient for the NC emission is opposite to that of the HVC2: while there is a clear pattern of an increase in offset with velocity in the HVC2, the slower LVC NC emission is offset further from the star. The meaning of these features are discussed in Section 4.

\begin{figure*}
\centering
   \includegraphics[width=8cm, trim= 0cm 0cm 0cm 0cm, clip=true]{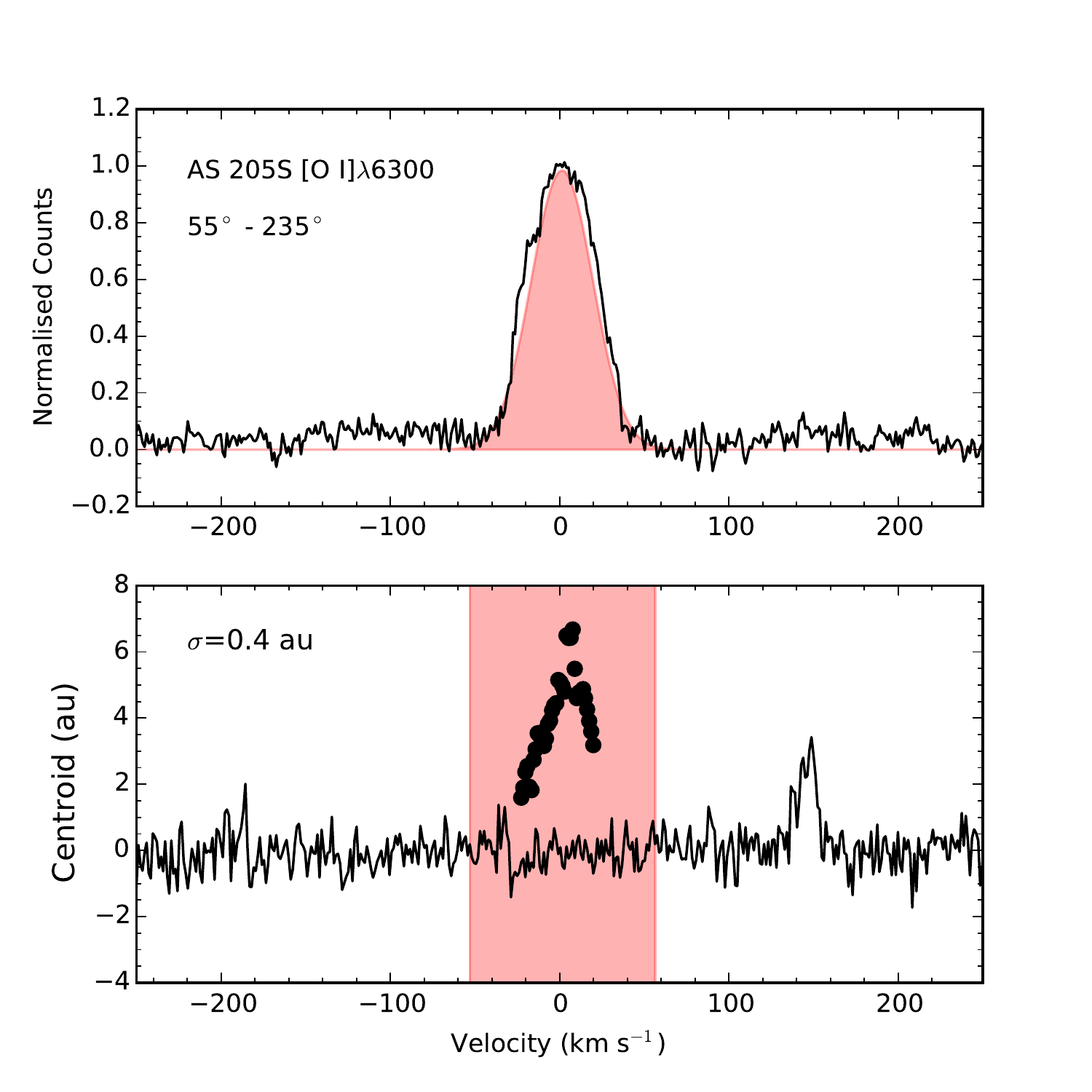}
    \caption{{\bf AS~205~S:} Spectro-astrometric analysis of the [O I]$\lambda$6300 emission. Top: The continuum subtracted line emission, normalised to the line peak. The fitted kinematic component is over-plotted. Bottom: The position spectrum with the range of the fitted LVC-BC shown as the colored shaded region. The continuum subtracted offsets are over-plotted in black.}
  \label{OIAS_S}     
\end{figure*}

\begin{figure*}
\centering
   \includegraphics[width=12cm, angle=-90, trim= 0cm 0cm 0cm 0cm, clip=true]{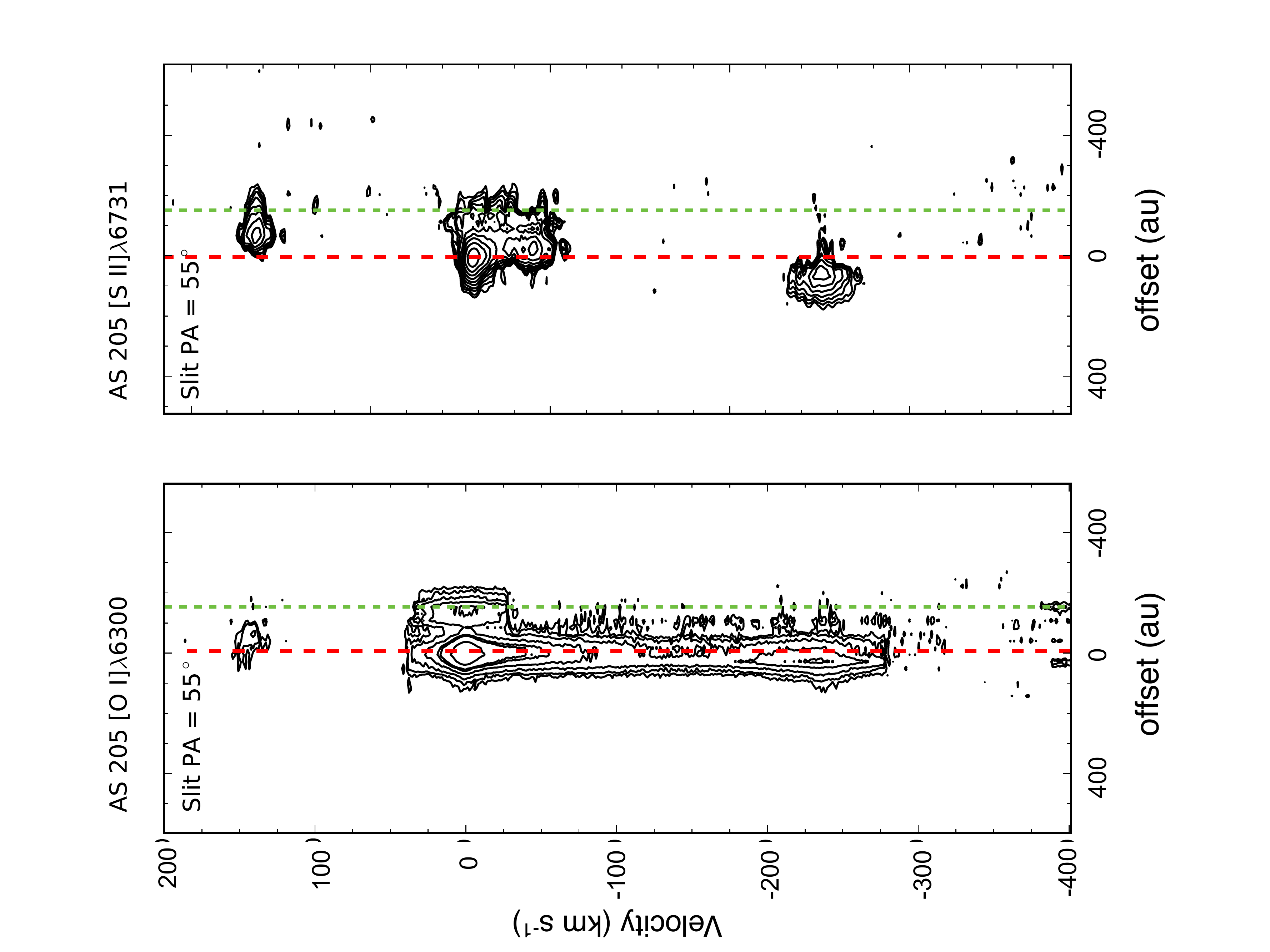}
    \caption{Continuum subtracted position velocity diagram of the \OIa\ and \SIIa\ emission from AS~205 for the 55$^{\circ}$ slit PA. Both AS~205~N and AS~205~S were included in the slit and the position of AS~205~S is marked with the green dashed line.}
  \label{PV}     
\end{figure*}


\subsection{AS~205: A multiple System} \label{sec:res_as205}

As described in Section~\ref{sect:targets} the AS~205 system is a multiple system with three known members. The aim here was to investigate the low velocity emission from AS~205~N. However, for the 55$^{\circ}$ and 235$^{\circ}$ slit PAs AS~205~S~a,b are also included in the UVES slit.  Therefore, the emission from AS~205~S could also be considered (keeping in mind that As~205~S a,b is not resolved). Considering the disk PA for AS~205~N of 114$^{\circ}$ and the binary PA of 212$^{\circ}$, a complication with interpreting any spectro-astrometric signal becomes apparent. As any outflow would be expected to be launched approximately perpendicular to the accretion disk, and as the binary PA is also approximately perpendicular to the AS~205~N accretion disk, it will be difficult to distinguish a binary signal from an outflow signal. However, it would be reasonable to assume that any forbidden emission is originating in an outflow.

For AS~205~N both blue and red-shifted displacements are detected in the forbidden emission lines. Again the \OIb\ emission region is not found to be displaced and the 1-$\sigma$ uncertainty in the SA can again be taken as an upper limit on the extent of this emission region. For AS~205~N this is 2~au (after continuum subtraction) which equals 6~au when the inclination of the system is accounted for. Analysis of the \OIa\ and \SIIa\ line regions reveals interesting results.  As presented in Figure \ref{RULUP_full} the HVC3 and the HVC1 are displaced in the same direction while the HVC2 is displaced in the opposite direction. As HVC1 and HVC2 are both blue-shifted one would expect them to be extended in the same direction, if they are part of the same flow. The complex nature of the signals detected becomes clear in Figure \ref{AS205N_CONT_2D}, where the continuum subtracted results for perpendicular slits are combined to recover the PA of the signals. The four identified components are shown separately. The magenta line here marks the PA of AS~205~S with respect to AS~205~N. The red line gives the PA of the AS~205~N disk major axis. In \OIa\ HVC2, HVC3 and HVC1, while offset in different directions, all lie along similar PA.  The same directions are seen for the three HVCs in \SIIa. Also note that the PAs of the extended components are very close to the binary PA and are also close to being perpendicular to the disk PA.  

The analysis of the AS~205~S \OIa\ line is presented in Figure \ref{OIAS_S}. No \SIIa\ emission was detected for AS~205~S. No offset is measured for the \OIa\ before continuum subtraction but there is a signal at the velocity of the red-shifted emission detected for AS~205~N and in the direction of AS~205~N. After continuum subtraction this red-shifted feature is located at $\sim$ 100~au from AS~205~S, in agreement with its location with respect to AS~205~N and the separation of $\sim$ 163~au between the two stars. Also the AS~205~S \OIa\ emission line shifts in the direction of AS~205~N after continuum subtraction. In Figure \ref{PV}, position velocity diagrams of the \OIa\ and \SIIa\ emission line regions are presented offering a further picture of the system. The location of the red-shifted emission feature between the two sources is clear.  

As noted, it is difficult to distinguish between a binary and outflow signal in this system. However, it is argued here that a likely explanation is that we are detecting an asymmetric jet (V$_{blue}$ $\sim$ -250~kms$^{-1}$ and V$_{red}$ $\sim$ 140~kms$^{-1}$) driven by AS~205~N, with the signals seen in HVC1, the \SIIa\ LVC, and the AS~205~S \OIa\ emission being binary signals. The red-shifted jet emission is located beyond the shrouding effects of the disk. High angular resolution imaging in jet tracing lines could help to fully understand the outflow activity in the vicinity of the stars. We conclude that the nature of the LVC remains elusive until the flows can be disentangled and driving sources identified.

\begin{figure*}
\centering
   \includegraphics[width=16cm, angle=0, trim= 0cm 6.5cm 0cm 5cm, clip=true]{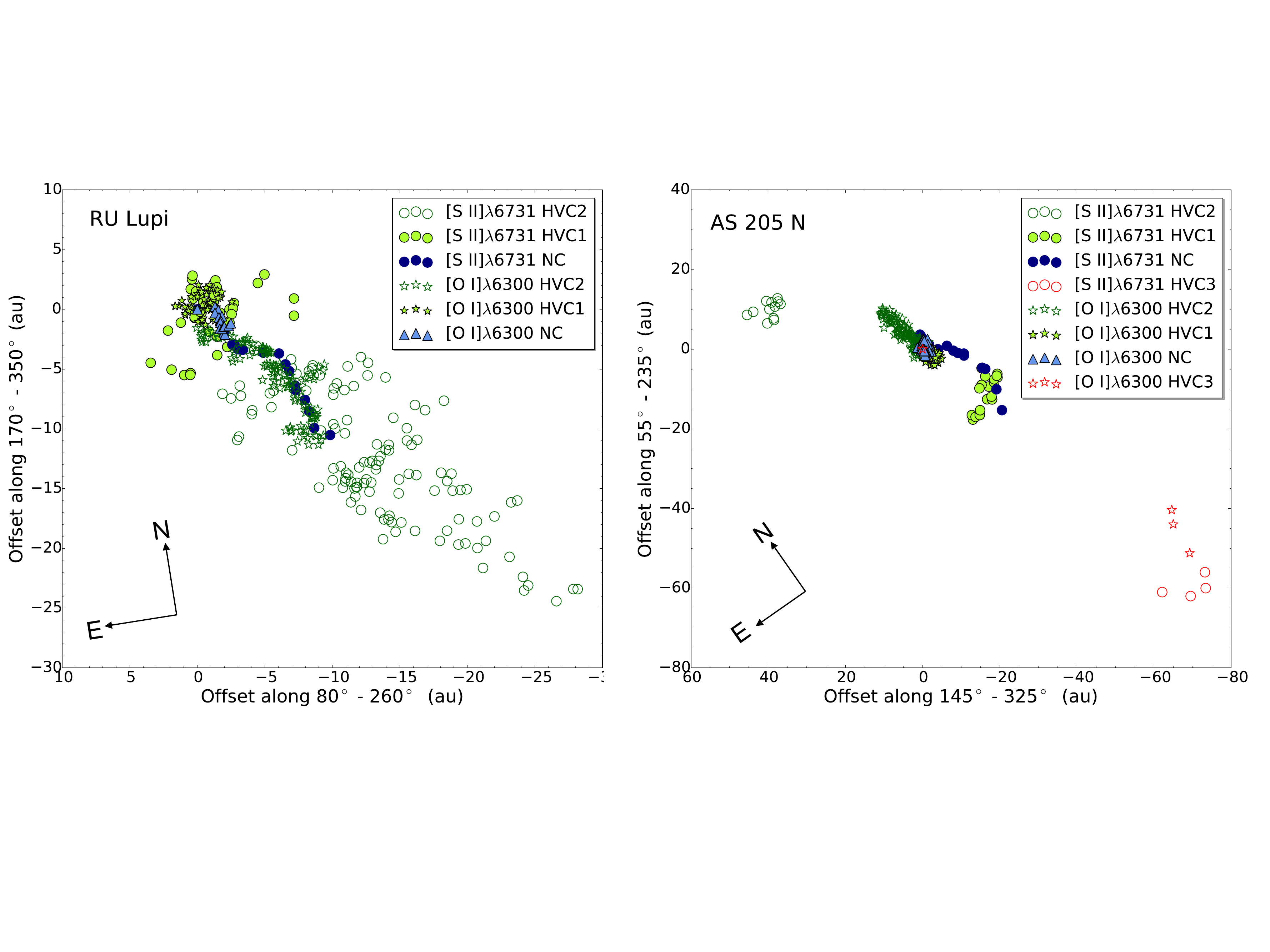}
     \caption{Comparing the spatial properties of the different kinematical components. For the blue-shifted components the HVCs are shown in green to represent their larger velocity. The black stars show the stellar positions. The agreement in the PA of the different RU~Lupi components is clear while in the case of AS~205~N the \SIIa\ NC and HVC1 do not agree with the idea of there being one outflow, as they are clearly in the direction of the red-shifted flow.  }
  \label{composite}     
\end{figure*}

\begin{figure}
\centering
   \includegraphics[width=12cm, angle=-90, trim= 0cm 0cm 0cm 0cm, clip=true]{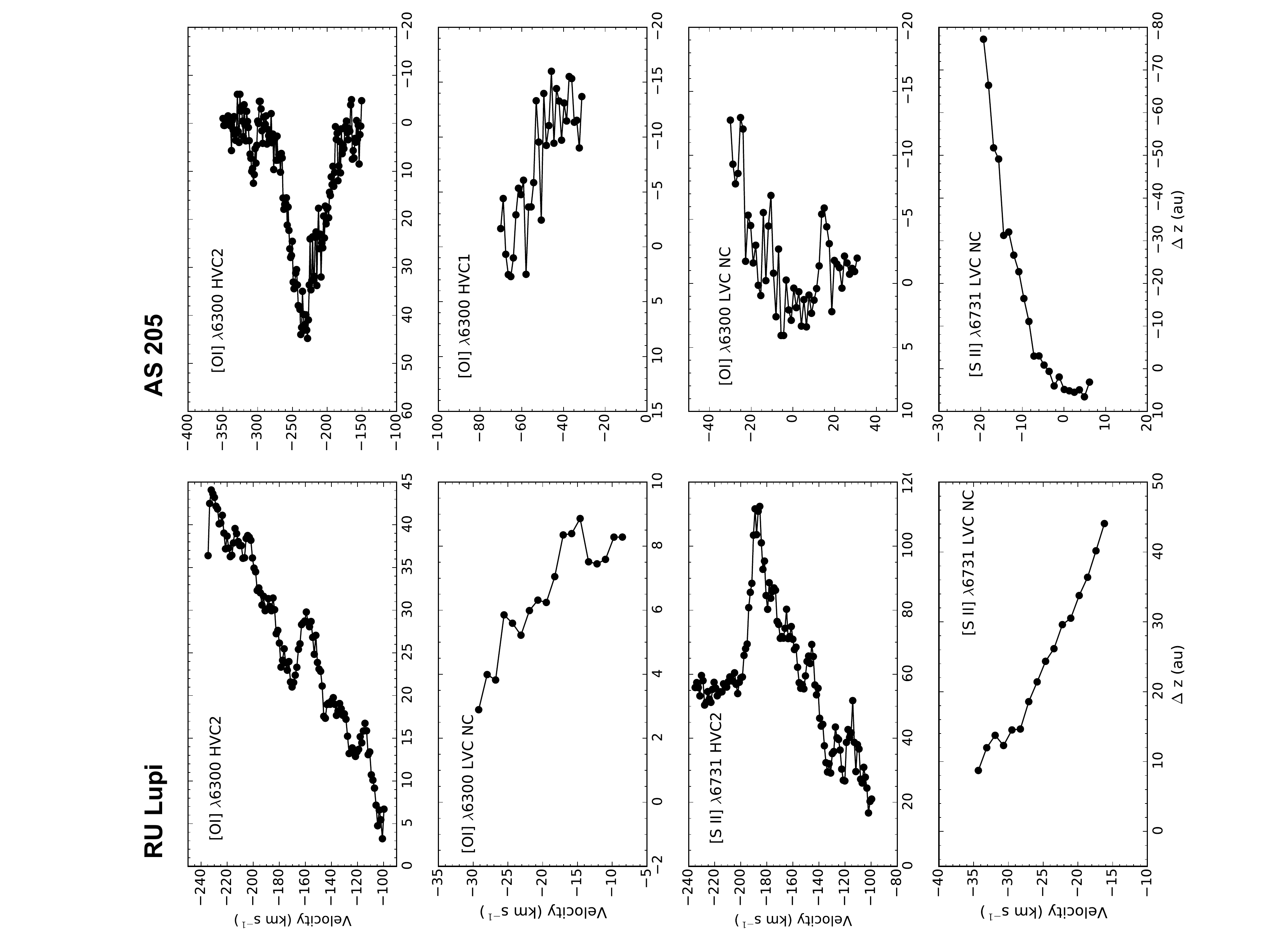}
     \caption{De-projected spatial offset along outflow PA, as a function of radial velocity for the RU~Lupi and AS~205~N HVCs and HVCs and LVC NCs. These results were extracted from Figures 2 and 4 are the average of the results for the two sets of perpendicular slit PAs. {\bf RU~Lupi:} While the jet shows a general trend of an initial increase in velocity with distance (typical of TT jets) a negative gradient is seen in the LVC NC emission. {\bf AS~205~N:} For all components except the \OIa\ HVC1 an initial increase in displacement with velocity is seen. This could be due to mixing between the HVC1 and HVC2, which are displaced in opposite directions.}
  \label{gradientRULup}     
\end{figure}



\begin{figure}
\centering
\includegraphics[width=12cm, angle=-90, trim= 0cm 0cm 0cm 0cm, clip=true]{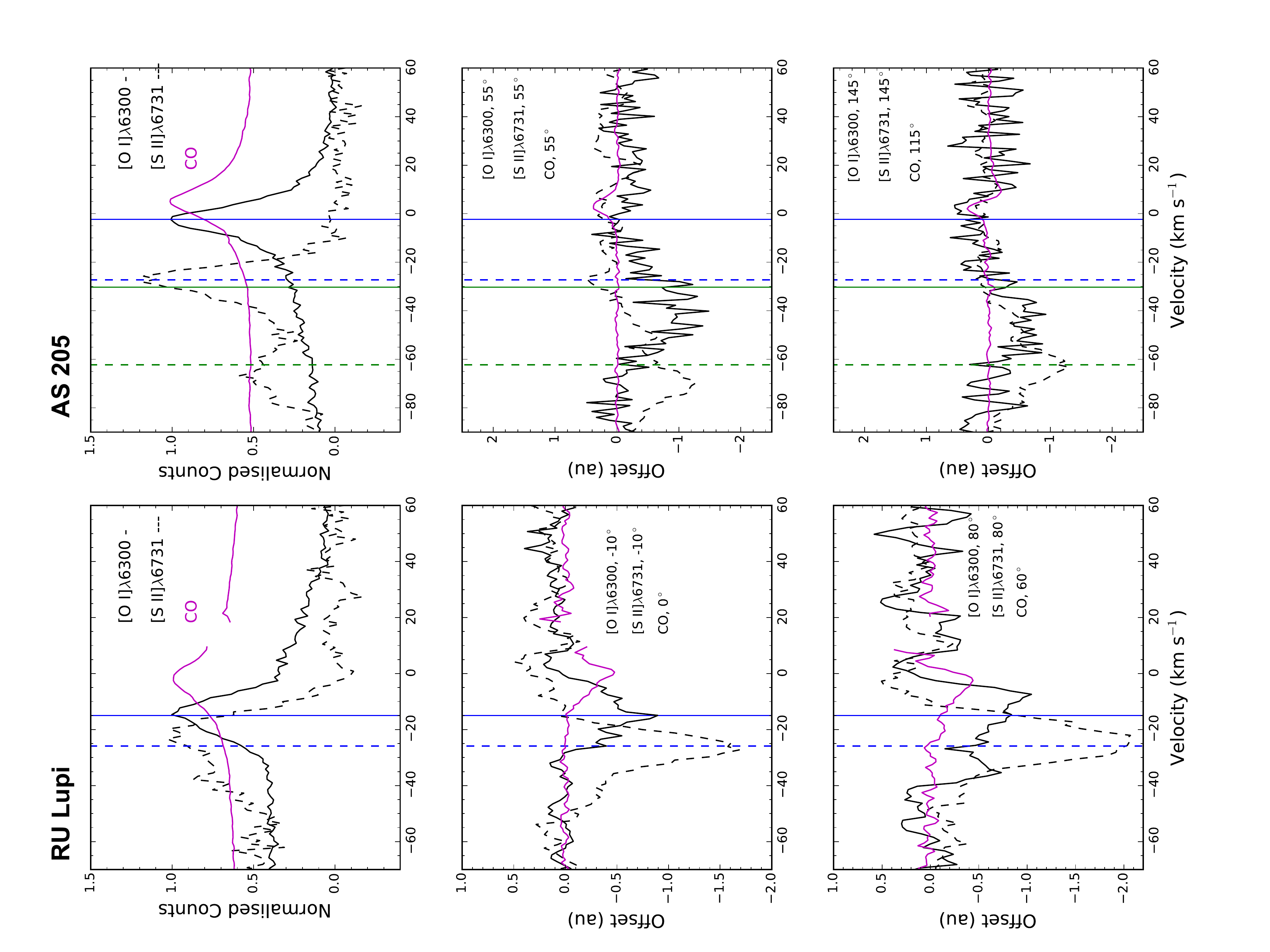}
  \caption{Comparison of the RU Lupi and AS~205~N spectro-astrometric results for the LVC NC to what was found for the CO by \cite{Pontoppidan2011}. The CRIRES CO spectra and the UVES spectra were taken at different slit PAs, so the spectra closest in PA are compared. Also the CO results were not corrected for continuum subtraction and so what is compared is the offsets measured before continuum subtraction and shown in Figures 1 and 3. The vertical lines mark the velocities of the forbidden line NCs and HVC1s in both sources. They are shown dashed for the [SII]$\lambda$6731 line. {\bf RU~Lupi:} The shape of the offsets is similar between the two lines. The maximum offset is at the peak of the line and then offset gradually tails off as velocity increases. This shows that the LVC NC of the FELs and CO have the same origin. {\bf AS~205~N:} Unlike what is found for RU~Lupi there is no agreement between the FEL and CO results here emphasising that an MHD disk wind is not detected in this case.}
  \label{RULUP_ponto}     
\end{figure}


\section{Discussion} 

As outlined above, spectro-astrometry of visible spectra of RU~Lupi and AS~205~N was previously presented by \cite{Takami2001, Takami2003}. What is new about the analysis presented here is that the much improved spectral and spatial resolution (better seeing conditions) offered by observing with UVES, means that the different kinematical components which make up the forbidden emission can be easily distinguished and analysed separately. Furthermore, the improved spectral resolution allows the relationship between displacement and velocity to be probed  at good resolution. In the case of AS~205 the improved seeing allows the N and S components to be separated. Another important development since the analysis by \cite{Takami2001, Takami2003} is the spectro-astrometric analysis of the CO emission from both stars by \cite{Pontoppidan2011}, who conclude that the CO fundamental at $\sim$4.7\,\micron{} is tracing a disk wind. 


While the kinematic analysis of the RU~Lupi and AS~205~N forbidden line emission summarised in Table \ref{fits} could suggest a similar spatial origin (within each component), the spectro-astrometric analysis presented in Figures 2 and 4 clearly points to a different origin. The important information to be taken from these figures are the PAs of the extended emission and the relationship between velocity and displacement. From Figure 2, it is clear that the HVC2 (tracing the jet) and the LVC NC have the same PA but for AS~205~N (Figure 4) the fastest component assumed to trace the jet (HVC2) and the displacement seen in the \SIIa\ LVC NC are in opposite directions. 
The agreement in the PA of the different RU~Lupi components is further highlighted in Figure 5. Here the different components are over-plotted without the kinematical information given in Figures 2 and 4. The NCs are shown with filled blue symbols and the HVCs with unfilled green and red (AS~205~N) symbols. Green is chosen for the blue-shifted HVCs to reflect the higher velocities with respect to the NCs. The agreement in PA for RU~Lupi is striking while for AS~205~N the \SIIa\ NC and HVC1 are not straightforward to explain.

 The distribution of the offsets with velocity also offer a clue to the origin of the emission. For the RU~Lupi HVC2 there is a clear initial increase in velocity with displacement along the disk rotation axis, as would be expected for a collimated jet, while for the LVC NC velocity decreases with increasing vertical offset. This negative velocity gradient is seen in both \OIa\ and \SIIa\ emission lines. Due to the complicated nature of the signals seen in AS~205~N the velocity distribution is less clear in Figure 4,  although the HVC2 in \OIa\ and the LVC NC in \SIIa\ do show a positive gradient (an increase in offset with velocity), while the HVC1 shows a negative gradient in the \OIa. To further clarify how the velocities are distributed with distance for the LVC NCs, the spatial offsets along the outflow PAs, as a function of velocity, are plotted in Figure 6. In Figures 2 and 4 the offsets shown were not de-projected but they are in Figure 6. The plotted offset is therefore the height above the circumstellar disk ({\bf z}). 
 
 Observations of TT jets forbidden emission line regions have clearly shown that there is an initial increase in the velocity of the jet with distance from the driving source and this behaviour was first recorded by \cite{Hirth1997}. Decreases in velocity at individual knots in the jets are also observed highlighting the complicated knotty structure of the jets. \cite{Takami2001} note a similar behaviour in the H$\alpha$ and FEL regions of RU~Lupi. They point out that this would be an expected feature of magnetically driven flows and use this as supporting evidence for the origin of the extended H$\alpha$ emission in the RU~Lupi outflow. This jet-like behaviour is found here for the HVC2 component in RU~Lupi, and we conclude that this is also found for all the AS 205 N components except the HVC1 component in \OIa.

The RU Lupi LVC NC is clearly showing something very different to what would be expected for a jet. Not only does it show a negative gradient in z but the gradient is much smoother and does not show evidence of regions of acceleration and deceleration as might be expected for shocked emission. In agreement with previous work, we find that a likely explanation of the velocity gradient is emission in an uncollimated MHD disk wind. The gradient seen in the RU~Lupi NC was hinted at in the \cite{Takami2001} results.  \cite{Takami2001} also investigated the velocity field of the RU~Lupi \OIa\ and \SIIa\ low velocity emission and Figure~\ref{gradientRULup} here can be directly compared to their Figure 3. Based on their data in Figure 3 of their article, \cite{Takami2001} conclude that the \OIa\ LVC does not show a gradient and argue that this is consistent with origin in a moderately collimated MHD disk wind. For the \SIIa\ line they note a slight positive gradient which they say is due to blending between the HVC and LVC. Therefore, they cannot draw any conclusions about the \SIIa\ LVC gradient. The spectra analysed in \cite{Takami2001} were taken with the RGO spectrograph on the AAT telescope at much poorer spectral resolution than what was achieved here and so while they see a hint of what we find here in the \OIa\ line it is likely that the \SIIa\ LVC was not resolved from the HVC. 

\begin{figure}
\centering
\includegraphics[width=10cm, trim= 0cm 0cm 0cm 0cm, clip=true]{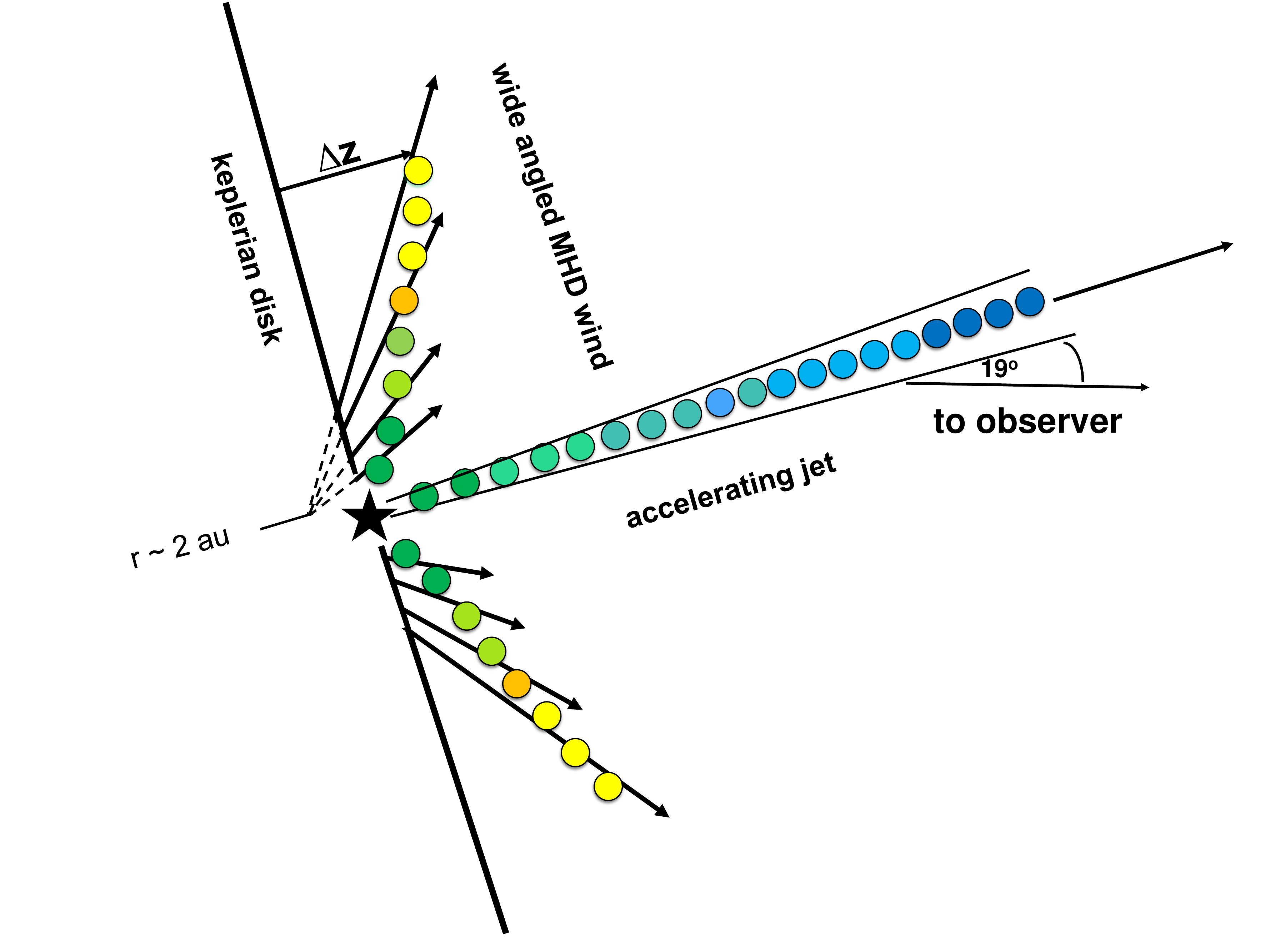}
  \caption{Sketch illustrating how the spectro-astrometric signals presented in Figures 1 and 2 are tracing the outflows from RU Lupi. The coloured dots are the positional displacements shown in Figure 2 with the colors from yellow (0 km/s) to dark blue (-230 km/s) representing increasing velocities. The HVC2 is tracing the accelerating jet with velocities from -90 km/s to -230 km/s. The LVC NC is tracing a wide angled MHD disk wind with a velocity range of -40 km/s to 0 km/s. For the MHD wind component, the velocity of each point corresponds to flow streamline and therefore a disk radius r. The spectro-astrometric technique is recording the height of the emission above the disk z, at each velocity (or r). As the velocity is decreasing with increasing r and z is increasing with r, the spectro-astrometric results shown in Figure 2 for the NC, has a negative velocity gradient.}
  \label{sketch}     
\end{figure}

The extent of the emission regions can also be deduced from Figure \ref{gradientRULup}. While the displacements shown in Figures 1 to 5 were corrected for the continuum contamination they did not take the inclination of the systems into account. In Figure \ref{gradientRULup} the displacements along the measured jet PA are plotted and they are also de-projected. For RU Lupi both the LVC NC and HVC2 are displaced further in \SIIa\ than in \OIa, as expected. From Figure \ref{gradientRULup} the RU~Lupi NC is displaced as far out as $\sim$ 8~au and $\sim$ 40~au in \OIa\ and \SIIa\ respectively, giving one of the first true measurements of the spatial extent of the LVC. The error on the measurement is given by the precision of the technique. The 1-$\sigma$ uncertainty is 0.6~au for \OIa\ and 1~au for \SIIa. 

Figure 7 compares the results from this work with the spectro-astrometric analysis of the CO emission from both stars presented by \cite{Pontoppidan2011}. For both sources the HVC1 and LVC NC components are delineated by vertical lines with the \SIIa\ components represented by a dashed line. The same color scheme as shown in previous figures is followed. We find that the RU~Lupi CO results agree with the FELs while there is no such agreement for AS~205~N. The comparison between the CO and FEL results for AS 205~N do show the signatures to be similarly shaped but then are along opposite PAs. 
\cite{Pontoppidan2011} model their CO spectro-astrometric results as being produced by a wide-angled wind and they also see the negative gradient detected here.

Since the RU Lup disk is at a low inclination angle, the line of sight velocity measures largely the velocity component along the disk rotation axis ($z$ direction).  A negative gradient in this velocity with $z$ therefore implies a decreasing $v_z$ and for a wind this is consistent with a decreasing flow velocity with distance from the star and/or with a decrease in the collimation angle of the flow streamline which also results in a decrease in the $v_z$ component of the flow velocity.  Both of these are predicted by MHD disk wind models. A similar fall off in velocity with disk radius has been observed in TT jets where an ``onion-like" velocity structure with fast flowing gas enclosed by slower less collimated gas, has been observed (see \cite{Frank2014} and references within).  This interpretation of the negative velocity gradient is also consistent with the conclusion of \cite{Fang2018} that while the BCs trace disk radii within 0.5~au and heights ($\sim$ 0.15~au) close to the disk (not detected in the UVES data), the NCs trace gas which originates
in a more extended wind at several au. Figure \ref{sketch} is a sketch of how properties of the outflows from RU~Lupi result in the spectro-astrometric offsets presented in Figures 1 and 2.



Our findings are very relevant to the ongoing debate about the origin of the LVC  emission. \cite{Simon2016} found that the combination of blue-shifts in the centroid and large FWHM of the LVC BC pointed to a MHD disk wind origin while the origin of the NC was much less clear and a photoevaporative wind could not be ruled out, see also \citet{EO2016}. As \cite{Banzatti2019} found that both the BC and NC kinematics were correlated, they further argued that they both are part of the same MHD wind and they put a constraint of 35$^{\circ}$ on the wind opening angle. While, \citet{Weber2020} claimed that these same correlations can be achieved in an MHD inner wind (producing the HVC and BC) plus an X-ray driven photoevaporative wind (giving rise to the NC), the predicted \SIIa\ line luminosities are orders of magnitude larger than observed, see discussion in \citet{Pascucci2020}, casting doubts on their proposed scenario. 

We note that while a declining flow velocity with disk radius and poor collimation is also expected with photoevaporation, a photoevaporative flow for the origin of the RU~Lupi NC is very unlikely.
Firstly, the observed blue-shifts of $\sim 30$\,km/s at low $z$, are too high to be consistent with photoevaporative flows which, instead, predict shifts at thermal speeds of 5~kms$^{-1}$ to 10~kms$^{-1}$  \citep{Ballabio2020}. Secondly, the [OI]$\lambda$6300 emission can be traced down to a vertical height z of  $\lesssim$2\,au from the disk midplane\footnote{The \OIa\ NC emission extends back closer to the star than the \SIIa\ emission but this is due to the lower critical density of the \SIIa\ line}, see Figure~8, with the wind being launched relatively deep within the gravitational potential well of the star.


We conclude, from the analysis of the velocity distribution of the different components and the comparison with the CO that the RU~Lupi NC is most likely tracing a MHD disk wind. \cite{Gangi2020} compare molecular Hydrogen emission at 2.12~$\mu$m with the \OIa\ emission from a sample of TTSs and conclude that the H$_{2}$ and \OIa\ LV NC traces the same wind but cannot distinguish between a photoevaporative and a MHD wind. We also note the recent ALMA observations detected a non-Keplerian envelope in $^{12}\mathrm{CO}$, extended from the RU~Lupi disk, which \cite{Huang2020}  comment could be the colder/radially extended counterpart of the inner MHD wind inferred in the CO fundamental line. The forbidden emission investigated here probes material closer to the fundamental CO emission, raising the interesting possibility of disk winds being present at nearly all disk radii. While it appears that the CO is tracing part of the blue-shifted outflow of AS~205~N the situation is uncertain for the FEL HVC1 and LVC NC. \cite{Salyk2014} also introduced some uncertainty as to whether the CO is tracing a disk wind in AS~205~N as they suggest that the CO spectro-astrometric signature could have been confused by the binary. Further investigation is need to disentangle the complicated outflow emission from the AS~205 system.  As a wide-angled wind would enclose the fast collimated jet \citep{Klassen2013}, it would be expected that the PA of the wind component would be found to be the same as the PA of the jet in AS~205, as seen for RU~Lupi.

\section{Summary and Conclusions}
We carried out UVES spectro-astrometry in the different kinematic components of the \OIa\ and [S II]$\lambda\lambda$6716, 6731 
from two accreting stars, RU~Lupi and AS~205~N.
For AS~205~N several outflow components were revealed indicating the presence of at least two outflows, which is not surprising given that it is a multiple system. Analysis of the relationship between velocity and offset for the different AS~205~N flow components generally showed them to be jet-like. For RU Lupi a clear jet component was detected  (HVC2, V$\sim$ $-  195$ km~s$^{-1}$) which showed an increase in velocity with distance and evidence of a knotty structure. Another HVC component (HVC1, V $\sim$ $-66$ km~s$^{-1}$) shows no offset and we argue that it traces the base of the jet. The most significant finding was for the RU~Lupi LVC NC which was found to be extended along the same PA as the jet (to a distance of $\sim$ 40~au in \SIIa) but with a negative velocity gradient. This is opposite to the behaviour seen for jets. The evidence extracted from the spectro-astrometric analysis for RU~Lupi and the comparison with a previous spectro-astrometric study of CO emission points to an origin in an disk wind for the RU~Lupi NC. This is the first clear evidence of spatially resolved emission in a MHD disk wind for a LVC NC in a FEL and one of the first measurements of the wind height. Furthermore, this work highlights that a kinematical study alone is not enough to clarify the origin of the different velocity components. From the kinematical analysis alone, the FEL regions of both stars seems similar, but spectro-astrometry tells a different story. This approach should now be extended to a greater sample of objects to see if the LVC NC routinely traces a disk wind or if indeed RU~Lupi is a unique object.

\acknowledgements{This work is based on data collected by UVES (089.C-0299) observations at the VLT on Cerro Paranal (Chile) which is
operated by the European Southern Observatory (ESO). We thank Klaus Pontoppidan for providing the CO position spectra originally published in \cite{Pontoppidan2011}. Emma Whelan would like to acknowledge support from The Maynooth University Seed Fund 2019. I.P., U. G., and S.E. acknowledge support from a Collaborative NSF
Astronomy \& Astrophysics Research grant (ID: 1715022,
ID:1713780, and ID:1714229). 
R.D.A.~acknowledges funding from the European Research Council (ERC) under the European Union's Horizon 2020 research and innovation programme (grant agreement No 681601).}

\appendix

\section{\OIa\ Variability}
\begin{figure*}
\centering
\includegraphics[width=18cm, trim= 0cm 0cm 0cm 0cm, clip=true]{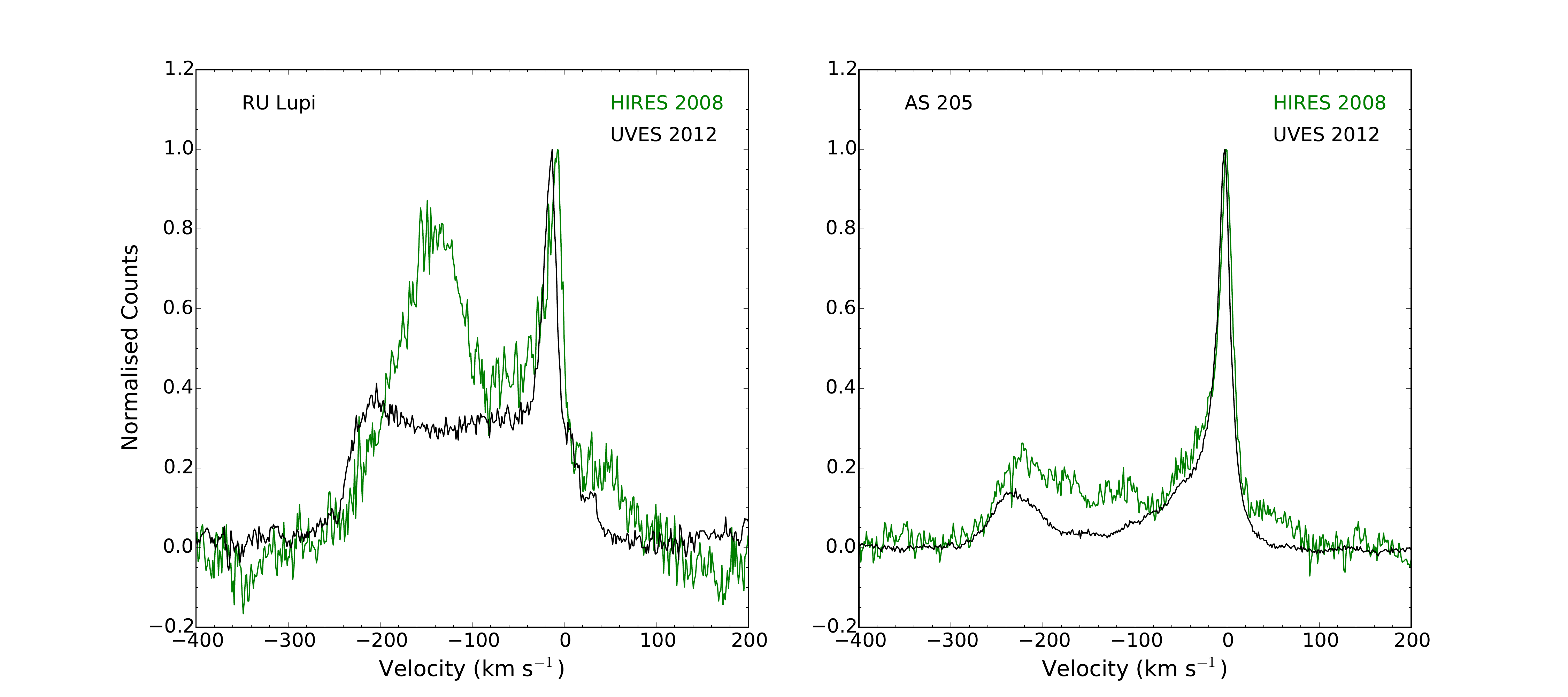}
  \caption{A comparison of the \OIa\ line profile in the HIRES (green) and UVES (black) spectra. In both cases the high velocity emission has decreased and this is most striking for RU~Lupi.}
  \label{OIcomp}     
\end{figure*}

The \OIa\ emission in RU~Lupi and AS~205~N were previously investigated in \cite{Banzatti2019} and \cite{Fang2018}. Both studies examined the same datasets, that is Keck/HIRES archival spectra taken in 2008. In Table \ref{fitscomp} the results of the kinematical fitting in these works are compared to our results. Both works identify a LVC broad and narrow component and two blue-shifted HVCs for both sources, although for RU~Lup the centroid and FWHM for the BC and one HVC are significantly different (see Table~4)\footnote{Note that Banzatti et al. (2019) report in their tables only the most blueshifted HVC but complete fits are available here: http://distantearths.com/Ilaria/disk-winds/}"
\cite{Banzatti2019} outline the changes in their approach to the kinematical fitting which accounts for these differences. The main difference between the results of this paper and these previous works is that no LVC BCs are identified.  In Figure \ref{OIcomp} we compare the normalised HIRES and UVES \OIa\ line profiles with the UVES. Note that at the epoch of the UVES observations the high-velocity emission has changed in both sources. This, in combination with the complex \OIa\ profiles, leads us to identify different Gaussian components than those published. Most notably, following the 30\,km/s separation for the LVC and HVC (e.g., Simon et al. 2016), we do not identify the weaker and more blended LVC BC in the UVES spectra. 
 
\begin{deluxetable*}{llll}
\tablecaption{Comparison of the \OIa\ decomposition from \cite{Fang2018} and \cite{Banzatti2019} to our own kinematic decomposition (last column).} 
\label{fitscomp}
\tablehead{ 
\colhead{}   &\colhead{Fang 2018} &\colhead{Banzatti 2019}   &\colhead{UVES} \\ 
\colhead{}   &\colhead{\small V$_{rad}$, FWHM (km~s$^{-1}$) } &\colhead{\small V$_{rad}$, FWHM (km~s$^{-1}$)}   &\colhead{\small V$_{rad}$, FWHM (km~s$^{-1}$)}}
\startdata
RU~Lup    & &  & \\
          &LVC NC (-10.9, 13.9) &LVC NC (-12, 15) &LVC NC (-13.0, 18.2) \\
            &LVC BC (-10.1, 176.4) &LVC BC (-36, 44) &not identified  \\
            &HVC (-32.3, 40.5) &HVC (-60, 292)  &HVC (-66.0, 150.0)  \\
            &HVC (-149.4, 109.9) &HVC (-148, 80) &HVC (-197.5, 59.5)  \\
AS~205~N   & &  & \\
           &LVC NC (-0.7, 14.5) & LVC NC (-0.4, 14.5) &LVC NC(-2.3, 16.8) \\
            &LVC BC (-13.5, 54.5) &LVC BC (-14.4, 54) &not identified  \\
            &HVC(-85.4, 225.5) &not identified &HVC (-30.3, 88.2)  \\
            &HVC (-221.9, 56.3) &HVC (-223, 58) &HVC (-226.5, 76.4)  \\
            &not identified &not identified &HVC (139.5, 13.7)  \\
\enddata
\end{deluxetable*}

{}


\begin{thebibliography}{}

\bibitem[Armitage(2011)]{Armitage2011} Armitage, P.~J.\ 2011, \araa, 49, 195. doi:10.1146/annurev-astro-081710-102521

\bibitem[Bai \& Stone(2013)]{Bai2013} Bai, X.-N. \& Stone, J.~M.\ 2013, \apj, 769, 76. doi:10.1088/0004-637X/769/1/76

\bibitem[Bailey(1998)]{Bailey1998} Bailey, J.\ 1998, \mnras, 301, 161

\bibitem[Balbus \& Hawley(1991)]{Balbus1991} Balbus, S.~A. \& Hawley, J.~F.\ 1991, \apj, 376, 214. doi:10.1086/170270

\bibitem[Ballabio et al.(2020)]{Ballabio2020} Ballabio, G., Alexander, R.~D., \& Clarke, C.~J.\ 2020, \mnras, 496, 2932. doi:10.1093/mnras/staa1767

\bibitem[Banzatti et al.(2019)]{Banzatti2019} Banzatti, A., Pascucci, I., Edwards, S., et al.\ 2019, \apj, 870, 76

\bibitem[Brannigan et al.(2006)]{Brannigan2006} Brannigan, E., Takami, M., Chrysostomou, A., \& Bailey, J.\ 2006, \mnras, 367, 315 

\bibitem[Cahill et al.(2019)]{Cahill2019} Cahill, E., Whelan, E.~T., Hu{\'e}lamo, N., et al.\ 2019, \mnras, 484, 4315

\bibitem[Dekker et al.(2000)]{Dekker00} Dekker, H., D'Odorico, S., Kaufer, A., Delabre, B., \& Kotzlowski, H.\ 2000, \procspie, 4008, 534 

\bibitem[Edwards et al.(1987)]{Edwards1987} Edwards, S., Cabrit, S., Strom, S.~E., et al.\ 1987, \apj, 321, 473

\bibitem[Ercolano \& Owen(2016)]{EO2016} Ercolano, B. \& Owen, J. E. 2016, MNRAS, 460, 3472

\bibitem[Ercolano \& Pascucci(2017)]{EP2017} Ercolano, B. \& Pascucci, I. 2017, RSOS, 470114

\bibitem[Ercolano et al.(2017)]{Ercolano2017} Ercolano, B., Rosotti, G.~P., Picogna, G., et al.\ 2017, \mnras, 464, L95


\bibitem[Fang et al.(2018)]{Fang2018} Fang, M., Pascucci, I., Edwards, S., et al.\ 2018, \apj, 868, 28

\bibitem[Frank et al.(2014)]{Frank2014} Frank, A., Ray, T.~P., Cabrit, S., et al.\ 2014, Protostars and Planets VI, 451

\bibitem[Gaia Collaboration et al.(2018)]{GAIA2018} Gaia Collaboration, Brown, A.~G.~A., Vallenari, A., et al.\ 2018, \aap, 616, A1

\bibitem[Gangi et al.(2020)]{Gangi2020} Gangi, M., Nisini, B., Antoniucci, S., et al.\ 2020, \aap, 643, A32. doi:10.1051/0004-6361/202038534

\bibitem[Gressel \& Pessah(2015)]{Gressel2015} Gressel, O. \& Pessah, M.~E.\ 2015, \apj, 810, 59. doi:10.1088/0004-637X/810/1/59

\bibitem[Hartigan et al.(1995)]{Hartigan1995} Hartigan, P., Edwards, S., \& Ghandour, L.\ 1995, \apj, 452, 736. doi:10.1086/176344

\bibitem[Hirth et al.(1997)]{Hirth1997} Hirth, G.~A., Mundt, R., \& Solf, J.\ 1997, \aaps, 126, 437

\bibitem[Huang et al.(2018)]{Huang2018} Huang, J., Andrews, S.~M., Dullemond, C.~P., et al.\ 2018, \apjl, 869, L42

\bibitem[Huang et al.(2020)]{Huang2020} Huang, J., Andrews, S.~M., {\"O}berg, K.~I., et al.\ 2020, \apj, 898, 140. doi:10.3847/1538-4357/aba1e1

\bibitem[Hu{\'e}lamo et al.(2008)]{Huelamo2008} Hu{\'e}lamo, N., Figueira, P., Bonfils, X., et al.\ 2008, \aap, 489, L9

\bibitem[Klaassen et al.(2013)]{Klassen2013} Klaassen, P.~D., Juhasz, A., Mathews, G.~S., et al.\ 2013, \aap, 555, A73. doi:10.1051/0004-6361/201321129

\bibitem[Koenigl \& Ruden(1993)]{Koenigl1993} Koenigl, A. \& Ruden, S.~P.\ 1993, Protostars and Planets III, 641

\bibitem[Kurtovic et al.(2018)]{Kurtovic2018} Kurtovic, N.~T., P{\'e}rez, L.~M., Benisty, M., et al.\ 2018, \apjl, 869, L44




\bibitem[McCabe et al.(2006)]{McCabe2006} McCabe, C., Ghez, A.~M., Prato, L., et al.\ 2006, \apj, 636, 932

\bibitem[McGinnis et al.(2018)]{McGinnis2018} McGinnis, P., Dougados, C., Alencar, S. H. P. et al. 2018, A\&A, 620A, 87





\bibitem[Nisini et al.(2018)]{Nisini2018} Nisini, B., Antoniucci, S., Alcal{\'a}, J.~M., et al.\ 2018, \aap, 609, A87



\bibitem[Pascucci et al.(2020)]{Pascucci2020} Pascucci, I., Banzatti, A., Uma, G. et al. 2020, ApJ, submitted





\bibitem[Pontoppidan et al.(2011)]{Pontoppidan2011} Pontoppidan, K.~M., Blake, G.~A., \& Smette, A.\ 2011, \apj, 733, 84 


\bibitem[Rigliaco et al.(2013)]{Rigliaco2013} Rigliaco, E., Pascucci, I., Gorti, U., et al.\ 2013, \apj, 772, 60


\bibitem[Salyk et al.(2008)]{Salyk2008} Salyk, C., Pontoppidan, K.~M., Blake, G.~A., et al.\ 2008, \apjl, 676, L49

\bibitem[Salyk et al.(2014)]{Salyk2014} Salyk, C., Pontoppidan, K., Corder, S., et al.\ 2014, \apj, 792, 68




\bibitem[Simon et al.(2016)]{Simon2016} Simon, M.~N., Pascucci, I., Edwards, S., et al.\ 2016, \apj, 831, 169

\bibitem[Takami et al.(2001)]{Takami2001} Takami, M., Bailey, J., Gledhill, T.~M., et al.\ 2001, \mnras, 323, 177

\bibitem[Takami et al.(2003)]{Takami2003} Takami, M., Bailey, J., \& Chrysostomou, A.\ 2003, \aap, 397, 675




\bibitem[Weber et al.(2020)]{Weber2020} Weber, M. L., Ercolano, B., Picogna, G., Hartmann, L., Rodenkirch, P. J. 2020, MNRAS, 496, 223


\bibitem[Whelan \& Garcia(2008)]{Whelan2008} Whelan, E., \& Garcia, P.\ 2008, Jets from Young Stars II, 123




\bibitem[Whelan et al.(2015)]{Whelan2015} Whelan, E.~T., Hu{\'e}lamo, N., Alcal{\'a}, J.~M., et al.\ 2015, \aap, 579, A48 



\end{thebibliography}
\end{document}